# Dielectric microsphere coupled to a plasmonic nanowire: A self-assembled hybrid optical antenna

*Sunny Tiwari[1†], Chetna Taneja[1†], Vandana Sharma[1], Adarsh Bhaskar Vasista[1#], Diptabrata Paul[1] and G. V. Pavan Kumar[1,2*]*

[1]Sunny Tiwari, Chetna Taneja, Vandana Sharma, Adarsh Bhaskara Vasista, Diptabrata Paul and G. V. Pavan Kumar
Department of Physics, Indian Institute of Science Education and Research, Pune-411008, India

[#]Present address: Department of Physics and Astronomy, University of Exeter EX44QL, United Kingdom

[2]Dr. G. V. Pavan Kumar
Center for Energy Science, Indian Institute of Science Education and Research, Pune-411008, India

[†]Authors contibuted equally.

*E-mail: pavan@iiserpune.ac.in

**Abstract**

Hybrid mesoscale-structures that can combine dielectric optical resonances with plasmon-polaritons are of interest in chip-scale nano-optical communication and sensing. This experimental study shows how a fluorescent microsphere coupled to a silver nanowire can act as a remotely-excited optical antenna. To realize this architecture, self-assembly methodology is used to couple a fluorescent silica microsphere to a single silver nanowire. By exciting propagating surface plasmon polaritons at one end of the nanowire, remote excitation of the Stokes-shifted whispering gallery modes (WGMs) of the microsphere is achieved . The WGM-mediated fluorescence emission from the system is studied using Fourier plane optical microscopy, and the polar and azimuthal emission angles of the antenna are quantified. Interestingly, the thickness of the silver nanowires is shown to have direct ramifications on the angular emission pattern, thus providing a design parameter to tune antenna characteristics. Furthermore, by employing three-dimensional numerical simulations, electric near-fields of the gap-junction between the microsphere and the nanowire is mapped, and the modes of nanowire that couple to the microsphere is identified. This work provides a self-assembled optical antenna that combines dielectric optical resonances with propagating-plasmons and can be harnessed in hybrid nonlinear-nanophotonics and single-molecule remote sensing.

How to effectively couple dielectric micro/nanostructures to plasmonic nanosystems? This is an important question in the context of hybrid nanophotonics, where the constituent dielectric and metallic elements can collectively give rise to optical effects which are otherwise not possible by individual elements. In recent times, such couplings have been experimentally realized on various systems such as metal-integrated semiconductor nanowires,[1] metal coated dielectric structures,[2] microsphere coupled with metallic particles,[3] dielectric nanospheres placed near metallic film.[4,5] These hybrid structures have been utilized, to design single photon devices[6-8], optical antenna[9,10] and for enhanced spontaneous emission from molecules[11,12] [13]down to the level of single molecule with minimized ohmic losses.[14,15]

In this paper, we introduce a self-assembled, hybrid photonic structure: dye molecule-coated silica microsphere coupled to a plasmonic-silver nanowire. The coated microspheres facilitate whispering gallery mode (WGM) resonances and silver nanowire provides propagating plasmons. We combine these characteristics and study directional emission of remotely-excited WGM resonances.

A variety of WGM resonators such as dielectric microspheres,[16] disks,[17-18] toroids,[15] nanogaps[19] have been fabricated/synthesized, and their optical properties have been extensively studied[20-22]. Specifically, the high-quality factors provided by these microresonators have been harnessed for various applications including lasing[23], non-linear photonics [24-26] and (bio) chemical sensors[27]. On the other hand, metallic structures such as nanoparticles, thin films, nanowires facilitate surface plasmons with very small mode volumes.[28,29] Large electric field produced by the localized surface plasmons in metal nanoparticles have been harnessed for a range of studies such as to probe enhanced emission from molecules placed near the nanoparticle,[30] bioimaging at subwavelength scale,[31] strong coupling at room temperature[32] whereas propagating surface plasmon polaritons (SPPs) in quasi 1D or 2D structures have been used for waveguiding properties[28]. One such structure which supports SPPs is chemically prepared silver nanowire.[33] The waveguiding characteristics via SPPs in silver nanowire have been extensively studied.[34,35] The directionality in the radiative outcoupling is facilitated by the end of nanowire, and can also be mediated by scattering of SPPs by a defect, such as a particle placed in the vicinity of nanowire.[36-39] This outcoupling is of relevance in realizing optical antenna, where directional emission is an important requirement. To simultaneously utilize the high quality factor of dielectric resonators and the large electric field with small mode volume of plasmonic structures, a large number of the studies have focused on coupling microsphere WGMs to localized surface plasmons in nanoparticles.[40,41] In contrast with nanoparticles, 1D structures supporting SPPs have not been coupled with dielectric structures which can be used to remotely excite the WGMs of the microsphere.

With the motivation of coupling WGMs of a dielectric microsphere with a plasmonic structure with waveguiding property, we experimentally study a dye coated $SiO_2$ microsphere coupled to a silver nanowire. Single $SiO_2$ microsphere can be optically coupled to a single plasmonic silver nanowire via self-assembly methodology without the necessity of any cleanroom facility or lithography procedures.[42] Specifically, we demonstrate remote excitation of WGMs of the microsphere via nanowire SPPs, and study their spectral signature and wavevector distribution using Fourier plane imaging. We find that emission from the nanowire-microsphere junction is directional, thus exhibiting optical antenna effect. Interestingly, the directionality in the emission, for both polar angles (θ) and azimuthal angles (φ), is introduced by silver nanowire, whose thickness can be tuned to control the emission angles. To further understand the near field coupling between microsphere and silver nanowire, we performed 3D finite element simulations. We calculated the electric near-field at the junction and simulated the modes of nanowire interacting with the microsphere.

The schematic of experimental configuration and methodology for preparing silver nanowire-dielectric microsphere (AgNW-µS) junction is shown in **figure 1a** and 1b. Chemically synthesized AgNWs in ethanol solution were dropcasted on a glass substrate and were left to dry.[33] Over this sample, dye molecules coated µS were dropcasted. After evaporation of solvent, the sample contained capillary force assisted self-assembled µS coupled AgNW [42]. A high numerical aperture objective lens (0.95 NA, 100x) was used to focus laser at one end of NW for the generation of SPPs. SPPs outcouple as free photons from the distal end of NW and the AgNW-µS junction. Figure 1c shows scanning electron microscope image, dark field image and bright field images of AgNW-µS junctions. We have probed one such junction which was prepared using AgNW of thickness ~ 350 nm and a dielectric µS of diameter ~3 µm coated with Nile blue (NB) molecules as shown in figure 1d(i). Figure 1 d(ii) shows the propagation of SPPs along AgNW at 633 nm excitation wavelength which is near to the absorption maxima of NB molecules. We have maintained polarization along the long axis of AgNW for efficient generation of SPPs throughout our experiments. As can be seen in figure 1d(iii), near field of SPPs and the outcoupled photons at the junction excite the fluorescence emission of molecules coated on the µS.[29,43] This resulting broad fluorescence emission acts as a local source to generate the WGMs of the µS. See section S1 of supplementary information for experimental setup.

To probe the spectral signature of a remotely excited dye coated µS coupled to an AgNW, we spatially filtered the junction using pinhole, placed at the image plane and projected the emission to the spectrometer. The spectrum shows the fluorescence of the molecules coupled to the WGMs of the µS (**figure 2**). The WGMs of the µS ride over the fluorescence spectrum and thus we see relatively intense modes near the fluorescence maxima of molecular emission. The analyzed spectrum of emission from the junction shows that the majority of the emission from the junction was polarized transverse to AgNW. (see sections S2 and S3 in supplementary information for NB fluorescence spectrum and input and output polarisation resolved spectrum of emission from the junction). The modes of the µS are assigned using Mie theory and are termed as transverse electric (TE) and transverse magnetic (TM) modes depending on the polarization of the mode.[20] The mode numbers are labeled using three numbers *n*, *l* and *m*. *n* represents the radial quantum number which quantifies the number of intensity maxima along the radius of µS. *l* represents the orbital quantum number which is equal to half of the intensity maxima along the perimeter of the µS. *m* represents the azimuthal quantum number which is the projection of orbital quantum number on the quantization axis (see sections S4 and S5 of supplementary information for discussion on assigning Mie modes and field profile). Since WGMs of µS are very sensitive to the environment,[44] we numerically calculated WGMs of a µS of diameter 3µm in air and compared it with the experimentally observed modes of NB molecules coated µS on glass substrate or near an AgNW (inset in figure 2).[45] Placing the microsphere on the glass substrate or near a silver nanowire shifts the peak position associated with the modes. See section S6 of supplementary information for the plot showing redshifting of modes of µS when it is placed on glass or near a silver nanowire. Along with the change in the peak position, the full width at half maxima of the peaks also broadens because of the interaction of near field of the µS with the environment. We have also performed the experiments with other sizes of microsphere. For µS of diameter 2 µm and 1 µm, coated with dye molecules with an excitation maxima near 575 nm and 520 nm wavelength respectively, we used 532 nm wavelength laser for SPPs generation along an AgNW. Spectral signature of µS of diameter 1 µm and 2 µm placed on a glass substrate and near an AgNW has been shown in S7 and S8 of supplementary information.

After probing the spectral signatures, we perform Fourier plane imaging to study the emission wavevectors from the AgNW-µS junction. Emission from the spatially filtered junction was projected to the EMCCD placed at the conjugate Fourier plane of the objective lens formed using relay optics.[46] Fourier plane image of emission from the junction shows that the majority of the emission is directed perpendicular to the length of AgNW and covers a short range of angles (**figure 3a-c**). The full width at half maxima (defined as α) which relates to the distribution in the radial angles of emission in the $+k_x/k_0$ direction is 0.34. For quantitatively measuring the directionality of emission from the junction, we used forward to backward ratio (in dB) of fluorescence emission along $k_x/k_0$ is given by

$$\beta = 10 \left(log \frac{I_+}{I_-}\right) \quad (1)$$

where $I_+$ and $I_{-se}$ is the intensity of fluorescence emission along $+k_x/k_0$ and $-k_x/k_0$ respectively. The emission from the junction is directional with a gain of 3.87 dB. In contrast to this result, when a NB molecules coated isolated µS placed on glass is directly excited, the Fourier plane image of the emission shows isotropic emission. The emission wavevectors cover a relatively large range of angles (figure 3d-f). Comparing both the Fourier plane images in figure 3b and 3e, we infer that AgNW acts like a reflector [47] for the emission from the µS and confines the emission in a narrow range of radial angles (θ), thus aiding optical antenna effect[48]. To study the effect of direction of SPPs along AgNW, on the directionality of emission, SPPs were launched individually from both the ends of AgNW and Fourier plane imaging was performed on the emission from the junction. Since, the area of µS in contact with AgNW is constant for both the excitation, the far field emission wavevectors are independent of the direction of SPPs along the AgNW as shown in section S9 of supplementary information. Fourier plane imaging of emission from µS of diameter 2 µm and 1 µm placed near an AgNW is shown in sections S10 and S11 of supplementary information.

Since the reflectivity of the AgNW results in directional emission from the junction, we ask whether the thickness of AgNW can be used as a parameter to tune the angular distribution of emission from the junction. For this study, we varied AgNW thickness for a constant µS size and performed Fourier plane imaging of emission from the junction. **Figure 4a** shows the bright field image of the junction prepared by NB molecules coated µS of diameter ~3 µm and an AgNW with thickness ~ 150 nm. Fourier plane image of the emission from the junction shown in figure 4a spans a large range of angles both in terms of θ and φ (Figure 4b). At a defined radial angle θ = 51°, the intensity profile (figure 4c) along the red dotted line in figure 4b shows a large distribution of emission along the azimuthal angles. Upon increasing the thickness of AgNW to ~ 250 nm, Fourier plane image (figure 4e) shows that the emission is biased in the forward direction ($+k_x/k_0$) with a reduced azimuthal distribution (figure 4f). As AgNW thickness was further increased to ~ 350 nm, majority of the emission can be seen to be confined in the $+k_x/k_0$ direction with a minimal azimuthal angle distribution as shown in Figures 4g-i. For a constant AgNW thickness, by changing the µS dimension the wavevector distribution of emission from the junction can also be tuned. This, in turn, changes the angular distribution of emission and gain in the directionality. See S12 of supplementary information for a discussion on variation of α and β with different AgNW thickness and µS size.

To further understand the near field coupling between AgNW and dielectric µS, we perform 3D finite element method simulations in COMSOL 5.3. An AgNW of diameter 350 nm and length 5 µm, with pentagonal cross section was modeled on a glass substrate. A dielectric µS of diameter 1 µm is placed equidistant from AgNW ends. The separation between dielectric µS and AgNw was 4 nm from one of the vertex of AgNW. The 4 nm separation represents the coating of polyvinylpyrrolidone (PVP) on the AgNW.[33] We have used refractive index for silver from the reference[49]. For launching SPPs, one end of AgNW was excited using a plane wave of wavelength 633 nm with polarization along the long axis of AgNW (**figure 5a**). At an excitation wavelength of 633 nm, a 350 nm thick AgNW placed on a glass substrate supports two leaky modes and one bound mode.[36, 50] Bound mode has an electric field confined between AgNW and the glass substrate, whereas the leaky modes have majority of the electric field confined on the vertices which are not in contact with the substrate.

Thus, we deduce that the electric field of leaky modes of AgNW is interacting with the µS. As shown in Figure 5b, the field at the vertex where the µS is coupled to the nanowire is relatively more intense as compared to the other vertex. Thus, an intense electric field at the gap between the µS and the AgNW vertex will lead to enhanced fluorescence emission from the molecules at the junction which will further result in increased coupling of fluorescence to WGMs. Thick nanowires will support more intense field at the junction compared to thin nanowires as the number of leaky modes reduce with a decrease in AgNW thickness (see S13 and S14 for the field profile of modes with variation of AgNW thickness and excitation wavelength). An intense electric field at the junction and reflectivity associated with thick AgNW contributes to intense and directional WGMs from the AgNW-µS junction.

In conclusion, we have experimentally shown how a capillary-force assisted self-assembled process can be utilized to couple a dye-coated µS to a plasmonic AgNW. The realized hybrid photonic structure was shown to act like a remotely excitable optical antenna. The nanowire coupled with the dielectric microstructure facilitates SPPs for the remote excitation of WGMs and the reflectivity of nanowire gives the directionality to the WGMs wavevectors. We quantified the directional emission from the hybrid structure and found that the thickness of the AgNW and size of the µS can be harnessed as parameters to control the emission angles of the out-coupled WGMs. Given that our geometry facilitates both propagating optical component via SPPs and a localized component in the form WGMs resonance, we envisage its utility as a directional photonic out-couplers on a micro-chip. In this work, we have essentially explored the spontaneous emission process in the form of WGMs mediated fluorescence, but we anticipate that the realized dielectric-metallic hybrid structure can be harnessed for stimulated emission studies, where a hybrid optical antenna can boost the directivity of symmetric resonators. Such directional emitters may find applications in nonlinear hybrid nanophotonics, random lasing and nano-optical sensing.

**Acknowledgments**

This work was partially funded by Air Force Research Laboratory and DST Energy Science (SR/NM/TP-13/2016) grant . Authors thank Deepak K Sharma, Shailendra K Chaubey for fruitful discussions and Raffeeque (Science and media center, IISER Pune) for drawing the schematic. Authors thank Harshith Bachimanchi for helping in sample preparation. S.T. thanks Infosys foundation and C.T. thanks Inspire fellowship for financial support. GVPK acknowledges DST for Swarnajayanti fellowship.

Sunny Tiwari and Chetna Taneja contributed equally to this work.

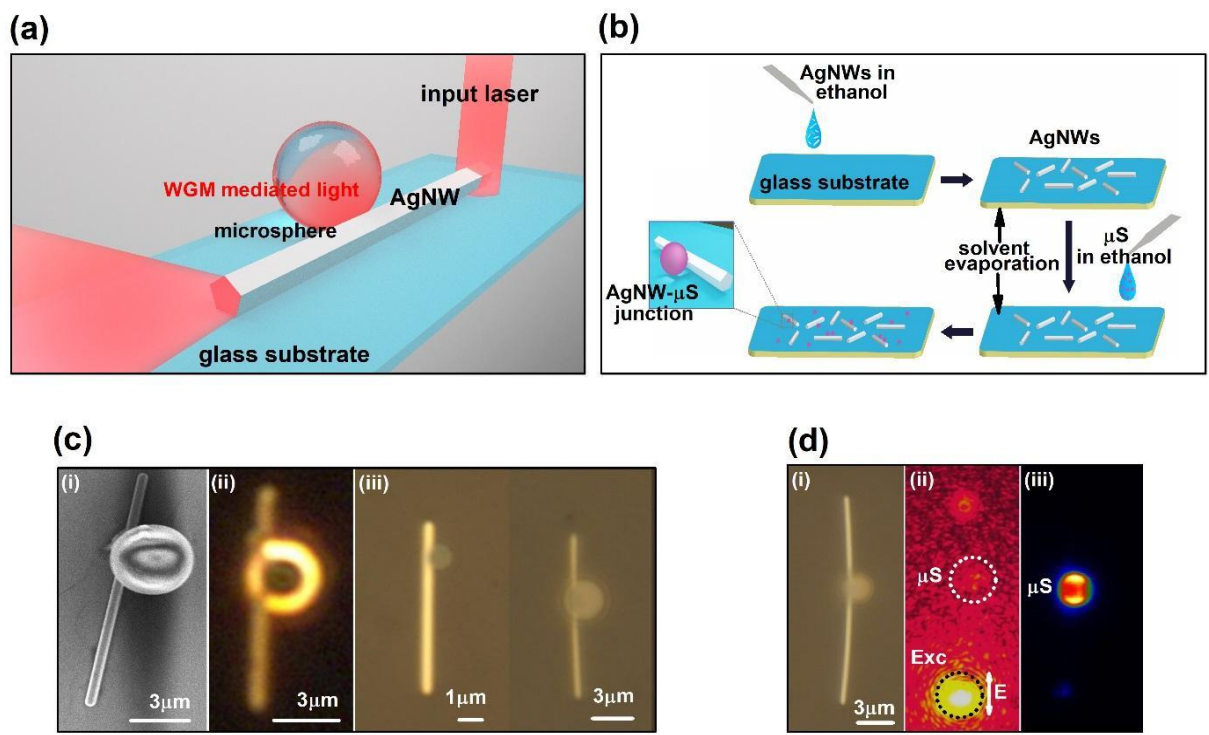

Figure 1. Schematic and imaging of the sample. a) Schematic of a silver nanowire (AgNW) coupled to a Nile blue molecules coated dielectric microsphere on a glass substrate. One end of AgNW is excited using a high numerical aperture lens and outcoupled light was collected from the silver nanowire-dielectric microsphere (AgNW-μS) junction. b) Methodology for preparing self-assembled AgNW-μS junction. c) (i) and (ii) Scanning electron microscope and dark field image of a ~ 3μm diameter microsphere coupled to ~ 350 nm thick silver nanowire respectively. (iii) Bright field images of AgNW-μS junctions with ~ 350 nm diameter thick AgNW coupled to a 1μm diameter μS and ~ 250 nm diameter thick AgNW coupled to a 3μm diameter microsphere. d) (i) Bright field image (ii) Elastic scattering image of the junction with ~ 350 nm wire and 3 μm sphere. (iii) Fluorescent image after rejecting the excitation wavelength, when plasmons along the nanowire are excited using a 633 nm laser.

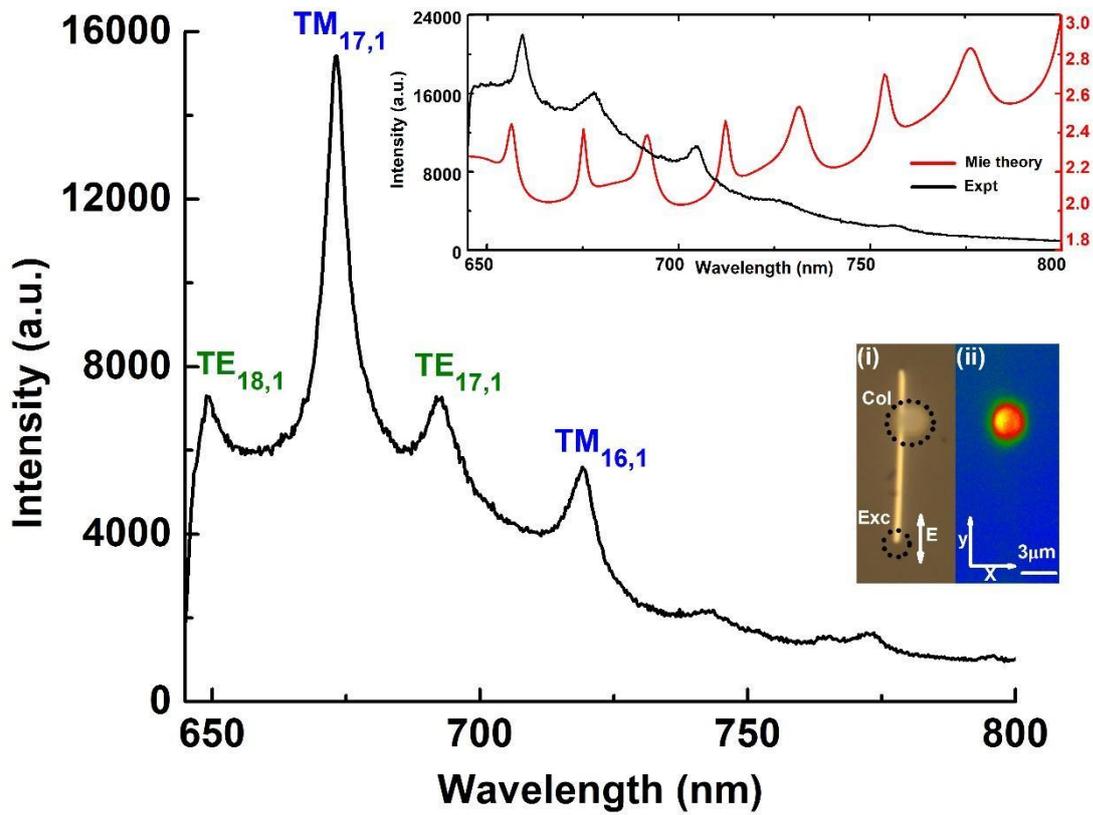

Figure 2. Spectral signatures of the hybrid photonic structure. We remotely excited whispering gallery modes (WGMs) of Nile blue coated-microsphere using surface plasmon polaritons of the nanowire. The spectrum shows sharp WGMs of a ~ 3 μm microsphere. Inset: (i) Bright field image. (ii) Fluorescent image of the junction. Plots in inset show calculated Mie resonances (TE/ $TM_{l,n}$) of an isolated microsphere of diameter 3 μm in air and experimentally measured spectrum of a microsphere of diameter ~ 3 μm placed on a graph substrate.

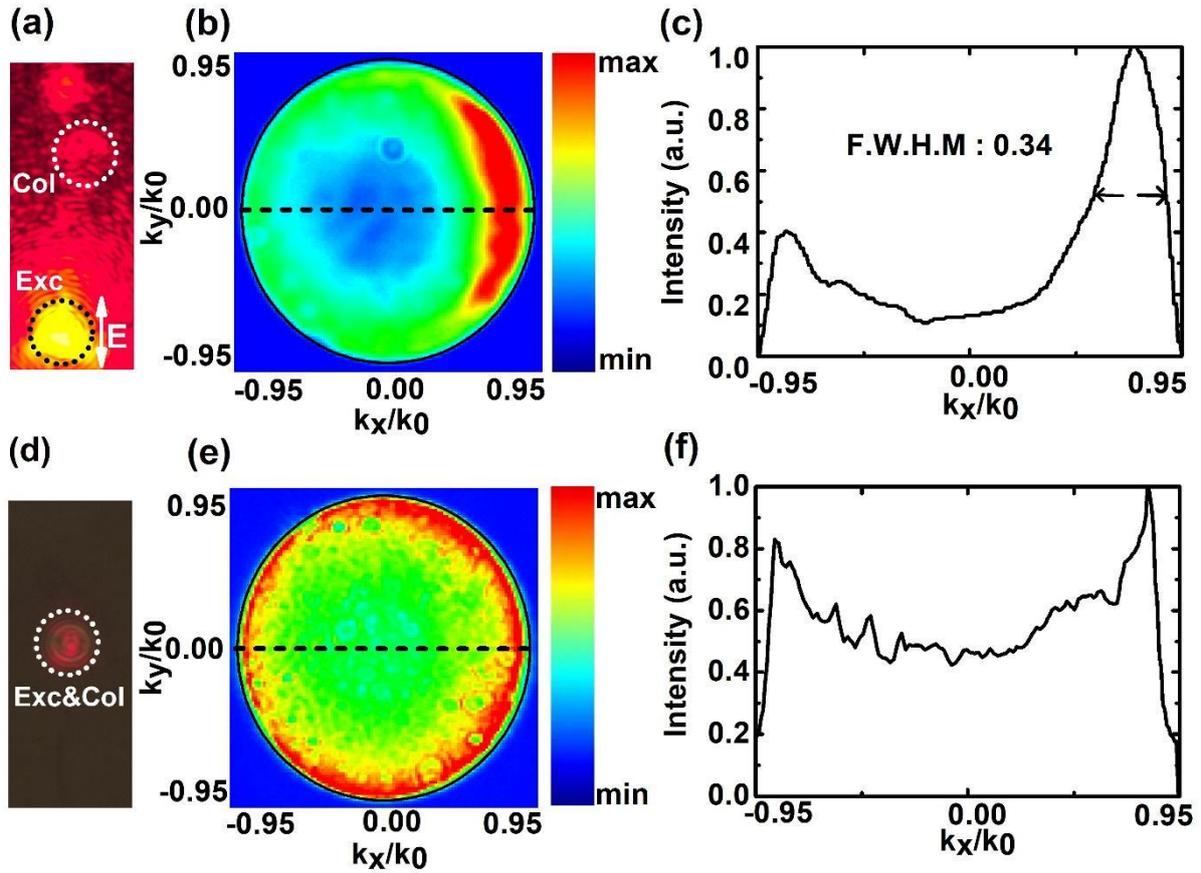

Figure 3. Comparison of the Fourier plane images of whispering gallery modes of remotely and directly excited dielectric microsphere. a) Elastic scattering image of remotely excited silver nanowire- dielectric microsphere (AgNW-μS) junction using a 633 nm laser. b) Fourier plane image of the spatially filtered, remotely excited fluorescence emission from AgNW-μS junction. Fourier plane image shows that the majority of the emission is confined to higher $+k_x/k_0$ values. c) Intensity profile (black dash line) along the $k_y/k_0 = 0$. The emission is confined to a short range of k vectors with a full width at half maxima (α) of 0.34 in the forward direction ($+k_x/k_0$). d) Elastic scattering image of a NB molecules coated microsphere on glass substrate. e) Fourier plane image of the fluorescence emission from the microsphere shown in d). f) Intensity profile (black dash line) along the $k_y/k_0 = 0$.

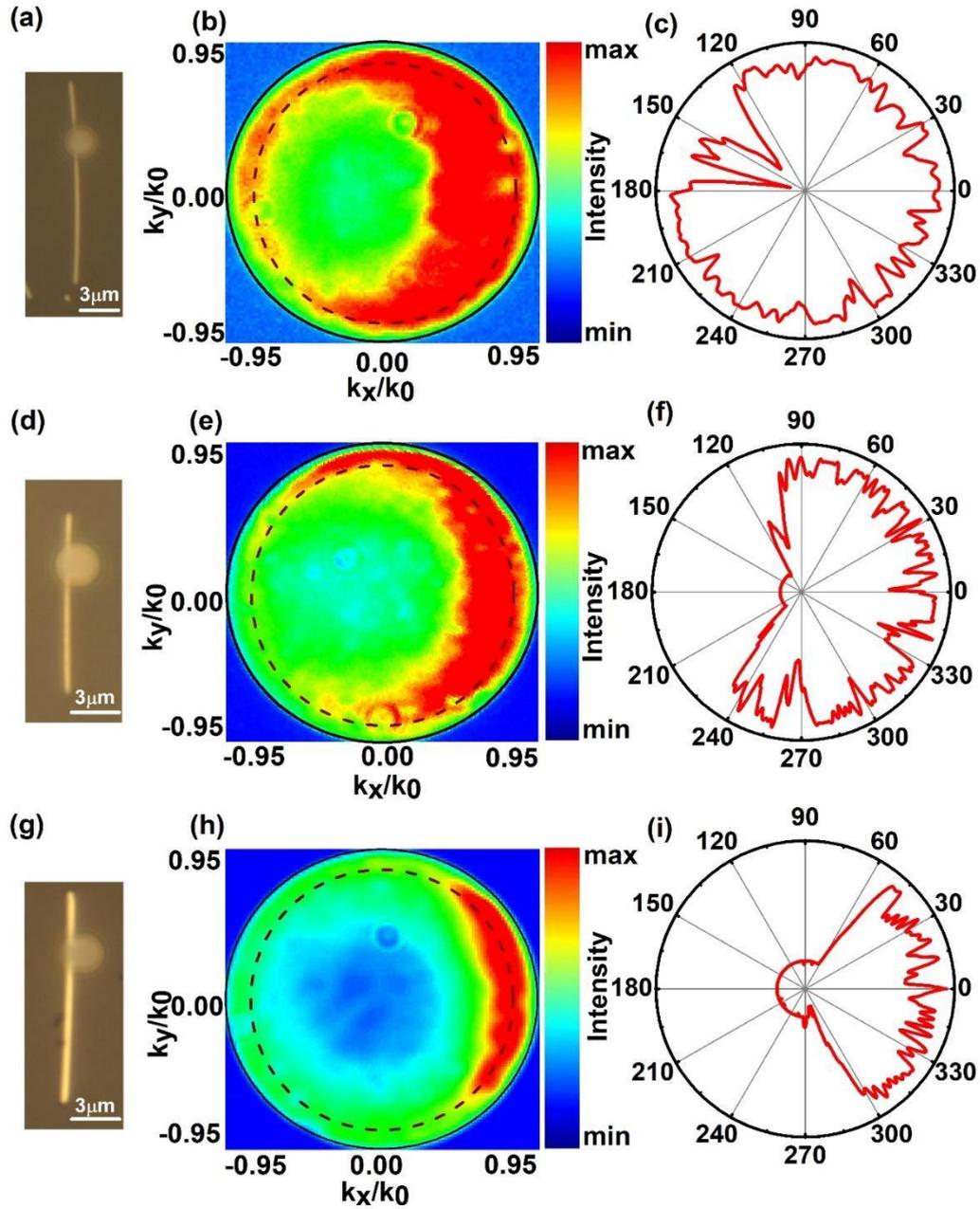

Figure 4. Controlling the azimuthal (φ) distribution of the emission by tuning the nanowire thickness. a) Bright field image of a ~ 3 μm microsphere coupled to a ~ 150 nm thick nanowire. b) Fourier plane image of emission from the junction spanning a large range of angular emission. c) The intensity profile of azimuthal angle along a defined radial angle, θ = 51° (black dotted curve in figure 1b). d)-f) Bright field image, respective Fourier plane image and intensity profile for a ~ 3 μm microsphere coupled to a nanowire of thickness ~ 250 nm. g)-i) Bright field

image, respective Fourier plane image and intensity profile for a ~ 3 μm microsphere coupled to a thick nanowire of diameter ~ 350 nm.

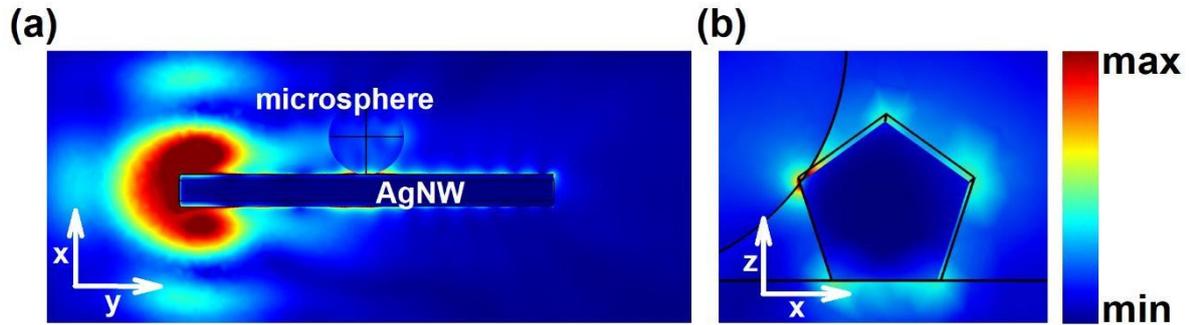

Figure 5. Finite element method simulations on silver nanowire on a glass substrate coupled to a dielectric microsphere. a) Electric near-field at the junction of a 1 μm diameter microsphere placed near a nanowire of thickness 350 nm on a glass substrate. Surface plasmon polaritons along the nanowire were launched by a 633 nm laser using a plane wave excitation with polarization along the longitudinal axis of nanowire. b) Electric near-field at the junction shows that field at one of the vertex of nanowire is interacting with the microsphere creating a relatively large local electric field as compared to the other vertices. The microsphere acts as a scatterer for the SPPs and leads to outcoupling of SPPs as free photons.

# Supporting Information
# Dielectric microsphere coupled to a plasmonic nanowire: A self-assembled hybrid optical antenna

*Sunny Tiwari[1], Chetna Taneja[1], Vandana Sharma[1], Adarsh Bhaskar Vasista[1,#], Diptabrata Paul[1], G. V. Pavan Kumar[1,2*]*

[1]Sunny Tiwari, Chetna Taneja, Vandana Sharma, Adarsh Bhaskara Vasista, Diptabrata Paul and G. V. Pavan Kumar
Department of Physics, Indian Institute of Science Education and Research, Pune-411008, India

[#]Present address: Department of Physics and Astronomy, University of Exeter EX44QL, United Kingdom

[2]Dr. G. V. Pavan Kumar
Center for Energy Science, Indian Institute of Science Education and Research, Pune-411008, India

*E-mail: pavan@iiserpune.ac.in

**Content:**
**S1.** Experimental setup
**S2.** Molecular fluorescence spectrum of Nile blue molecules dropcasted on a glass substrate
**S3.** Input and output polarization resolved spectrum of molecular emission from silver nanowire-dielectric microsphere junction
**S4.** Assignment of Mie modes for microsphere
**S5.** Calculated near field electric field distribution of a 3 μm dielectric microsphere placed on a glass substrate
**S6.** Red shifting of the modes of an isolated microsphere of size 3 μm with a change in effective refractive index of microsphere environment
**S7.** Spectral signatures of a dye coated 2 μm diameter microsphere
**S8.** Spectral signatures of a dye coated 1 μm diameter microsphere
**S9.** Effect of the direction of propagating plasmon polaritons along silver nanowire on the directionality of emission
**S10.** Fourier plane imaging of emission from a dye coated 2 μm diameter microsphere placed near a silver nanowire
**S11.** Fourier plane imaging of emission from a dye coated 1 um diameter microsphere placed near a silver nanowire
**S12.** Table for variation of full width at half maxima (α) of emission and forward to backward gain (β) of directionality in dB as a function of silver nanowire thickness and microsphere size
**S13.** Modes of silver nanowire of thickness 350 nm placed on a glass substrate at an excitation wavelength of 633 nm
**S14.** Modes of silver nanowire of thickness 150 nm and 250 nm placed on a glass substrate at an excitation wavelength of 633 nm and 532 nm

## S1. Experimental setup

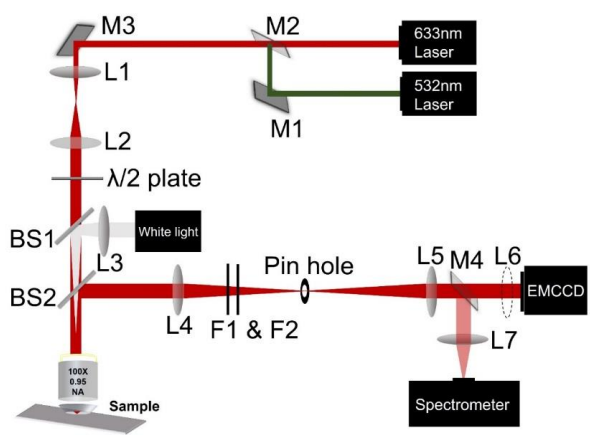

**Figure S1:** Experimental setup used to study the spectrum and wavevector signatures of a dielectric microsphere coupled to a silver nanowire.

    Figure S1 shows the experimental setup used to study the spectral and wavevector signatures of a dielectric microsphere coupled to a silver nanowire. The sample was excited using either 633 nm or 532 nm laser by a high numerical aperture objective lens (0.95, 100x) and the scattered light was collected using the same lens. M1 and M3 are mirrors and M2 is a flip mirror used to select either of the two lasers. Lenses L1 and L2 were used to expand the beam such that the back aperture of the objective lens was completely filled. The in-plane polarization of the laser was varied using a half wave plate ($\lambda/2$). The sample was observed using a white light source, lightly focused on the sample plane. BS1 and BS2 are beam splitters. L3, L4, L5, and L7 are lenses and L6 is a flip lens. M4 is flip mirror used to switch from collecting spectral signature to wavevector signature. Filters F1 and F2 are used to reject laser light and pinhole in the conjugate image plane is used to filter the nanowire-microsphere junction.

## S2. Molecular fluorescence spectrum of Nile blue molecules dropcasted on a glass substrate

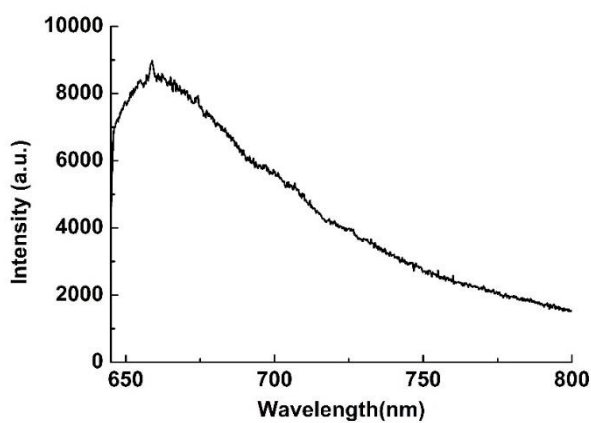

**Figure S2:** Molecular emission from a $10^{-5}$ Molar concentration of Nile blue molecules dropcasted on a glass substrate. The molecules in ethanol solution were dropcasted on a glass substrate and were let to dry. After evaporation of the solvent, molecules were excited using a 633 nm laser and the spectrum was collected after rejecting the laser light. The spectrum shows the fluorescence maxima of the molecular emission around 660 nm.

## S3. Input and output polarization resolved spectrum of molecular emission from silver nanowire-dielectric microsphere junction

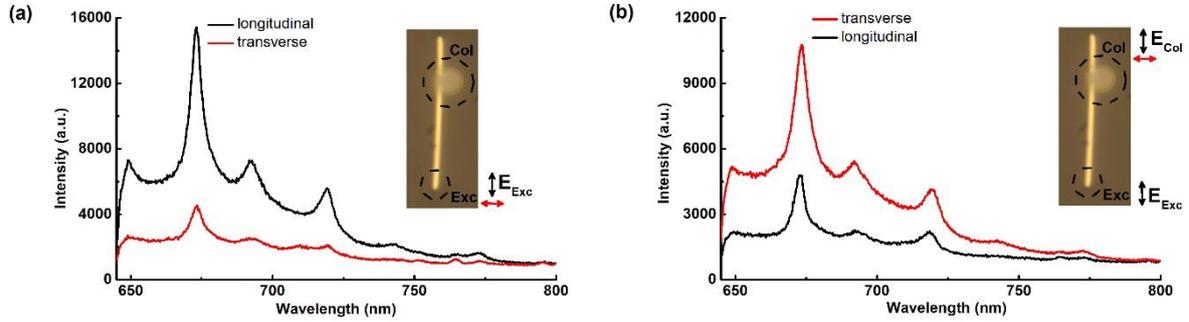

**Figure S3:** Input and output polarization resolved spectrum of molecular emission from silver nanowire-dielectric microsphere junction. a) Input polarization dependence of the spectrum from Nile blue coated microsphere. Silver nanowire when excited using a laser with polarization along the longitudinal axis results in efficient excitation of propagating surface plasmon polaritons (SPPs) as compared to when the polarization is kept perpendicular to the long axis. Thus with longitudinal input polarization, near field coupling of the SPPs with molecules increases and leads to more emission from the junction. b) Output polarization dependence. For output polarization dependence, input polarization is kept longitudinal to the nanowire. The Majority of emission from the junction is polarized transverse to the nanowire although the input polarization is longitudinal to the nanowire.

## S4. Assignment of Mie modes for microsphere

Whispering Gallery Modes (WGMs) are electromagnetic modes of a structure which depends on the morphology of the structures (here it is a microsphere). The waves suffer total internal reflection at the surface of the sphere and thus gets confined in a region of small volume. This creates a region of large local electric field.

We have calculated the modes of the microspheres using Mie theory. The codes for this calculation are given in reference [1].

The modes of a microsphere are labeled by two types of polarization, transverse electric (TE) and transverse magnetic (TM), and three quantum numbers n, l, and m. n is the radial quantum number which represents the number of intensity maxima along the radius of the sphere. l represents called the orbital quantum number which is equal to half of the intensity maxima along the perimeter of the microsphere. m which represents the azimuthal quantum number, is the projection of orbital quantum number on the quantization axis.

The extinction cross section for an interaction between a microsphere and electromagnetic field is given by,

$$\mathbf{Q_{ext}} = \frac{2}{(ka)^2} \sum_{n=1}^{\infty} (2n+1) Re(a_n + b_n)$$

Where, k is the wavenumber and a is the radius of the microsphere. Coefficients $a_n$ and $b_n$ are known as Mie coefficients. Putting $a_n = 0$ gives modes which are called transverse electric whereas setting $b_n = 0$ gives transverse magnetic modes.

## S5. Calculated near field electric field distribution of a 3 µm dielectric microsphere placed on a glass substrate

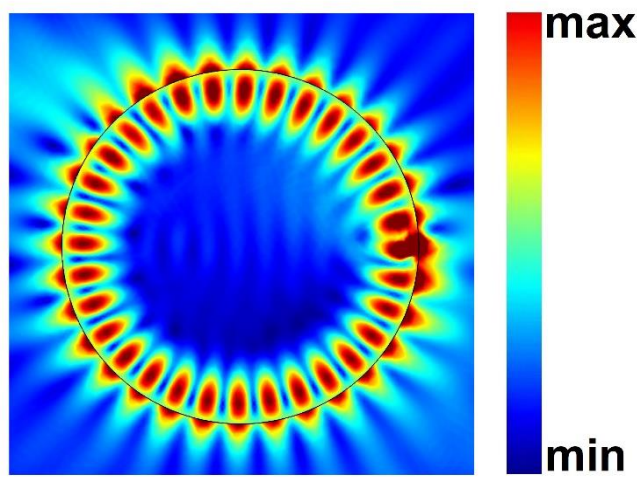

**Figure S5:** Calculated near field electric field distribution of a 3 µm dielectric microsphere placed on a glass substrate, using finite element method in COMSOL Multiphysics software. For excitation of the modes, we placed a dipole at the microsphere glass interface. The electric field is calculated at a wavelength of 656 nm which coincides with the position of a transverse magnetic (TM) mode with radial quantum number 'n' and orbital quantum number 'l' equals 17 and 1 respectively.

## S6. Red shifting of the modes of an isolated microsphere of size 3 µm with a change in effective refractive index of microsphere environment

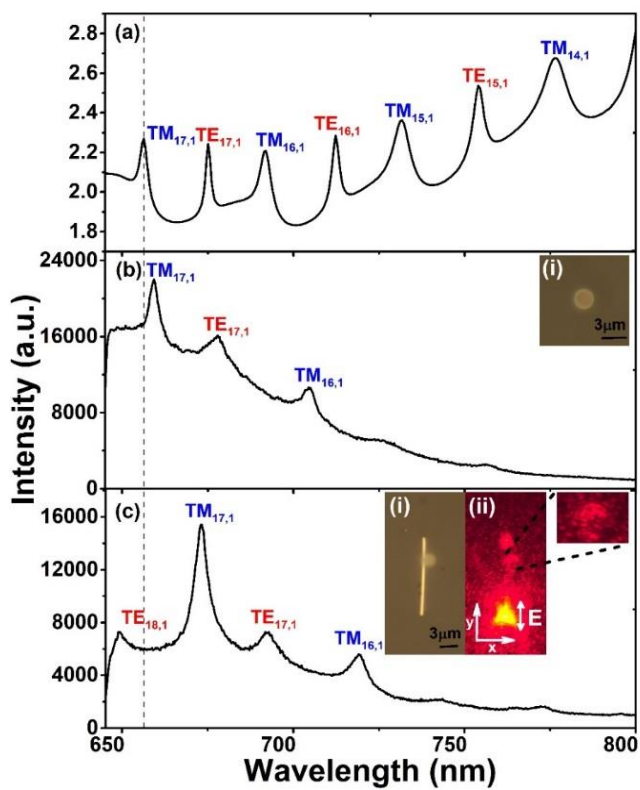

**Figure S6:** Red shifting of the modes of an isolated microsphere of size 3 µm with a change in effective refractive index of microsphere environment.

Figure S6 shows the red shifting of the modes of an isolated microsphere when it is placed on a glass substrate and near a silver nanowire. a) and b) Calculated spectrum for an isolated microsphere of diameter 3 µm using Mie theory and experimentally obtained spectrum for a microsphere of diameter 3 µm placed on a glass substrate respectively. Inset in (b) shows bright field image of a 3 µm diameter microsphere. c) Remotely excited modes of a microsphere of size 3 µm using SPPs along a silver nanowire of thickness ~ 350 nm. The black dotted vertical line originating from figure S5a) shows the wavelength position of $TM_{17,1}$ mode. Placing the microsphere on the glass substrate red shifts the mode $TM_{17,1}$ and also other modes. The shift is because of the change in the effective refractive index of the microsphere environment [2]. A larger red shift is observed when the microsphere is placed near a metallic nanowire.

**S7. Spectral signatures of a dye coated 2 *µ*m diameter microsphere**

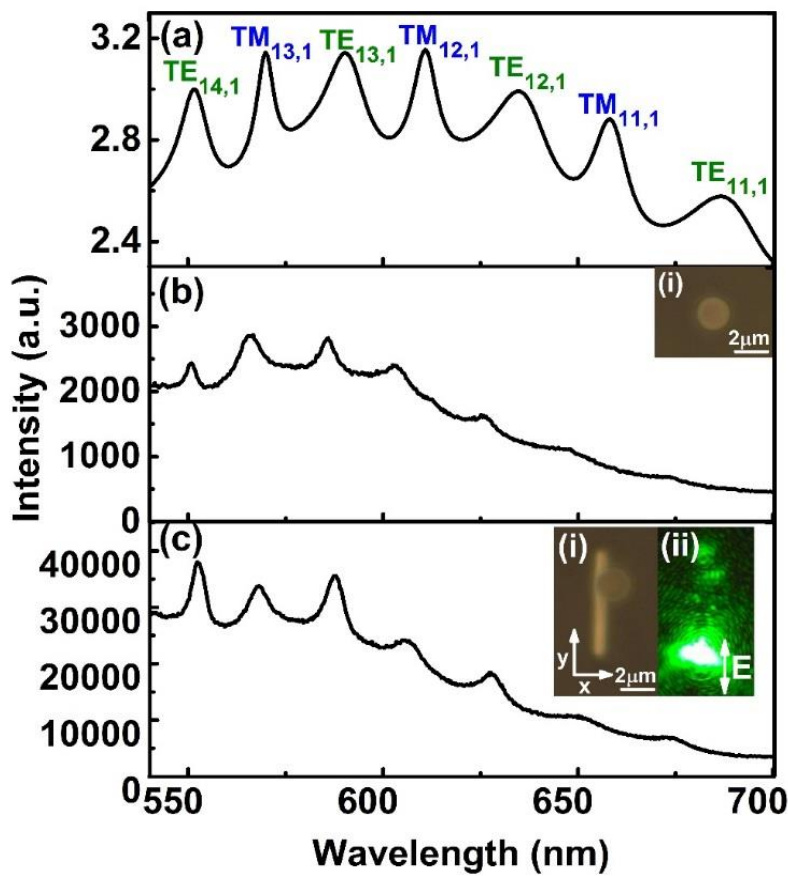

**Figure S7.** Spectral signatures of a dye coated 2 µm diameter microsphere placed near a silver nanowire of thickness ~ 350 nm. a) Calculated whispering gallery modes (WGMs) of an isolated microsphere of size 2 µm diameter using Mie theory. b) WGMs of a dye coated 2 µm diameter microsphere placed on a glass substrate excited using a 532 nm laser. The dye which are coated on the microsphere are resonant at a wavelength of 575 nm. Inset in b) shows a microsphere of size of 2 µm diameter placed on a glass substrate. Scale bar is 2 µm. c) Remotely excited WGMs of a dye coated 2 µm diameter placed near a silver nanowire of thickness ~ 350 nm. Propagating surface plasmon polaritons along the nanowire were launched by exciting one end of silver nanowire by a 532 nm laser. The junction is remotely filtered for collecting the spectrum after rejecting laser light as shown in inset (i) and (ii) of (c).

**S8. Spectral signatures of a dye coated 1 *μ*m diameter microsphere**

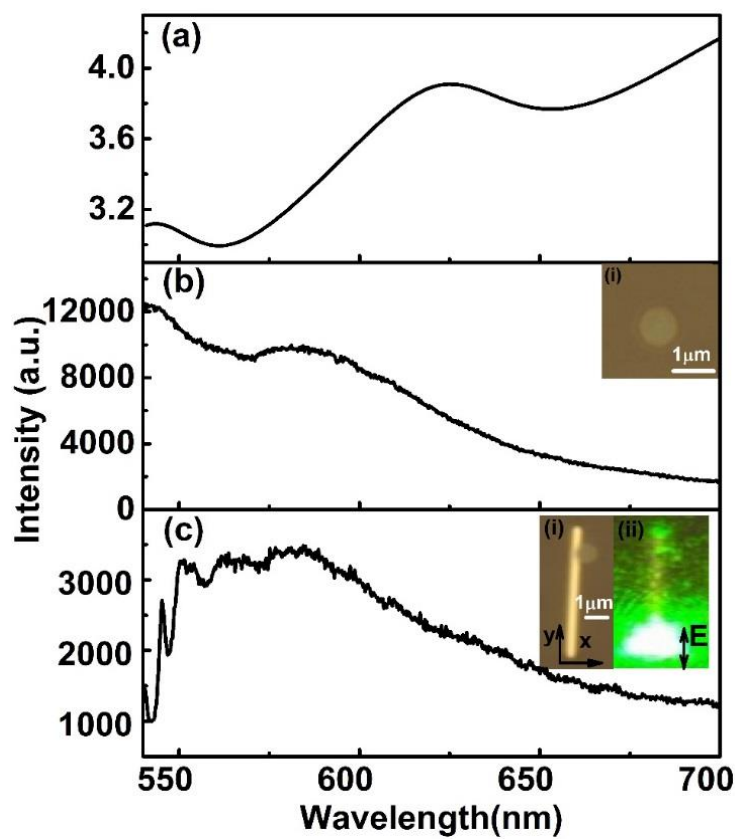

**Figure S8.** Spectral signatures of a dye coated 1 µm diameter microsphere placed near a silver nanowire of thickness ~ 350 nm. a) and b) Calculated modes of a 1 µm diameter microsphere using Mie theory and experimentally obtained modes of a dye coated microsphere of size 1 µm diameter placed on glass substrate, respectively. Inset in b) shows a 1 µm diameter microsphere placed on a glass substrate. Scale bar is 1 µm. (c) Remotely excited modes of a dye coated 1 µm diameter placed near a ~ 350 nm thick nanowire. The dye which are coated on the microsphere are resonant at a wavelength of 520 nm. Propagating surface plasmon polaritons along the nanowire were launched by exciting one end of silver nanowire by a 532 nm laser. The junction is remotely filtered for collecting the spectrum after rejecting laser light as shown in inset (i) and (ii) of (c). As can be seen in the calculated and experimentally obtained spectra, 1 µm diameter microsphere does not support WGMs in this range of wavelengths.

**S9. Effect of the direction of propagating plasmon polaritons along silver nanowire on the directionality of emission**

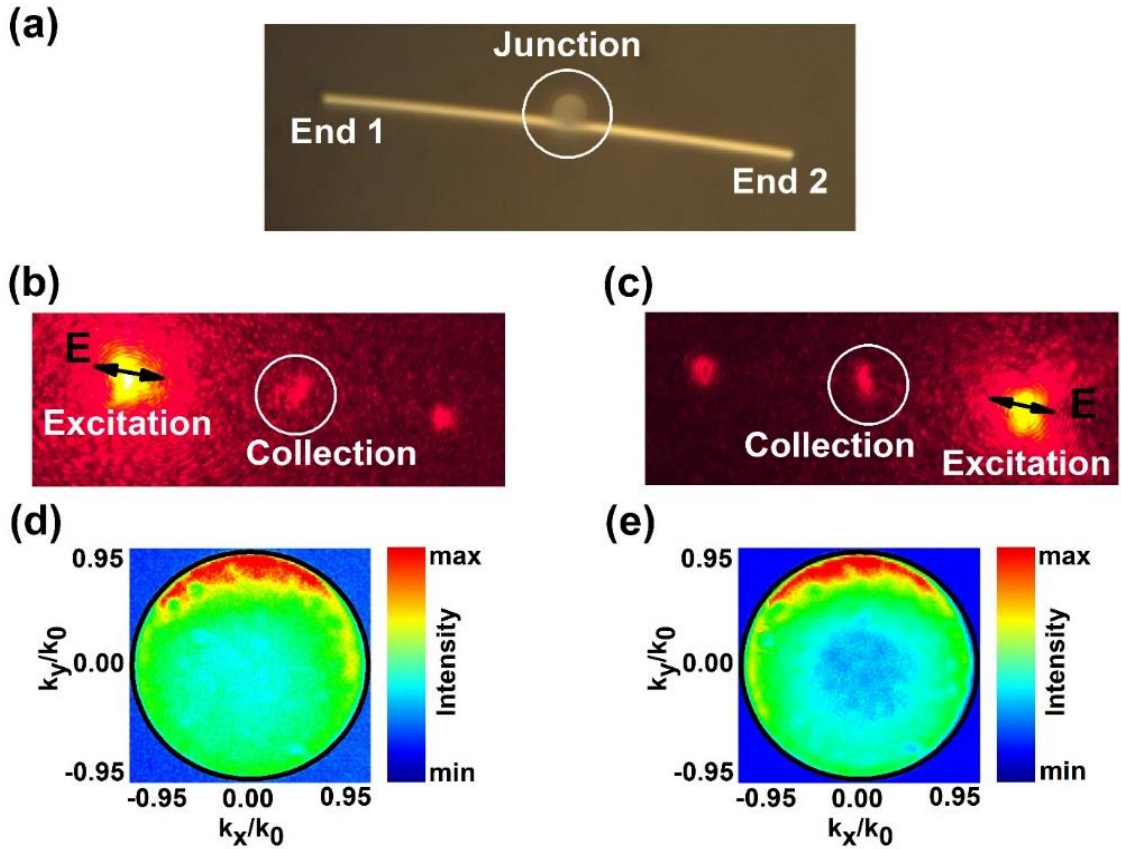

**Figure S9:** Effect of the direction of propagating surface plasmon polaritons along silver nanowire on the directionality of emission from junction.

Figure S9 shows the effect of propagating surface plasmon polaritons (SPPs) along the silver nanowire on the emission wavevector from the nanowire-microsphere junction. a) Nile blue coated microsphere was placed near a silver nanowire of thickness ~ 350 nm. b) and c) Position of laser excitation on the nanowire. SPPs along the silver nanowire were excited from end 1, propagates towards the end 2 in b) and vice-versa in c). d) and e) Fourier plane images of emission wavevectors from the junction when SPPs along the silver nanowire were excited from end 1 and end 2 respectively. The emission wavevectors from the junction is same in both the cases as the wire acts as a mirror and reflects the emission in same direction.

**S10. Fourier plane imaging of emission from a dye coated 2 *μ*m diameter microsphere placed near a silver nanowire**

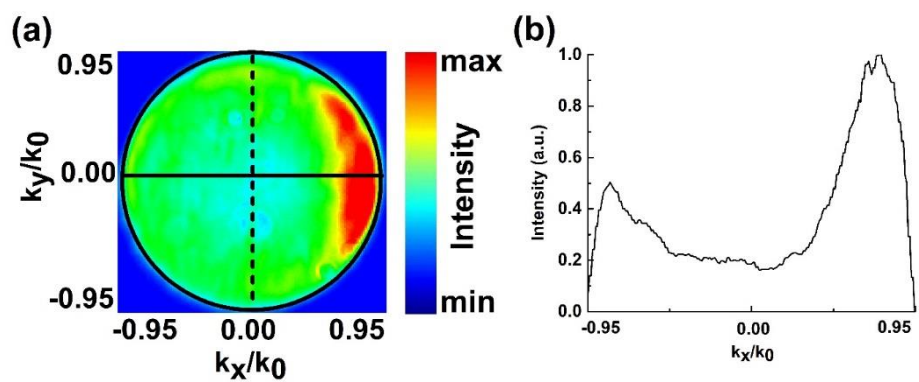

**Figure S10**. Fourier plane imaging of emission from a dye coated 2 µm diameter microsphere placed near a silver nanowire of thickness ~ 350 nm. Propagating surface plasmon polaritons along the nanowire are excited using a 532 nm laser. a) Emission wavevectors of molecular emission from silver nanowire-dielectric microsphere junction, with vertical dotted line as the longitudinal axis of the silver nanowire. b) Intensity profile along the black horizontal line in a). The emission from the junction is directed in a specific narrow range of wavevectors.

**S11. Fourier plane imaging of emission from a dye coated 1 *μ*m diameter microsphere placed near a silver nanowire**

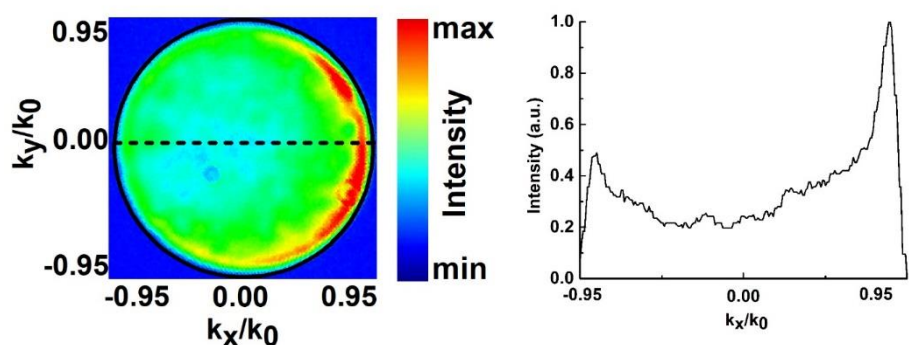

**Figure S11.** Fourier plane imaging of a dye coated 1 µm diameter microsphere placed near a silver nanowire of thickness ~ 350 nm. Propagating surface plasmon polaritons along the nanowire are excited using a 532 nm laser. a) Emission wavevectors of molecular emission from silver nanowire- dielectric microsphere junction, with vertical dotted line as the longitudinal axis of the silver nanowire. b) Intensity profile along the black horizontal line in a). The emission from the junction is directed in a specific narrow range of wavevectors.

## S12. Table for variation of full width at half maxima (α) of emission and forward to backward gain (β) of directionality in dB as a function of silver nanowire thickness and microsphere size

Variation of Full Width at Half Maxima (α) of emission and forward to backward gain (β) in dB as a function of silver nanowire (AgNW) thickness and microsphere size.

| AgNW thickness | microsphere size | | |
|---|---|---|---|
| | 3 μm | 2 μm | 1 μm |
| 350 nm | α = 0.34, β = 3.87 | α = 0.41, β = 3.01 | α = 0.18, β = 3.10 |
| 250 nm | α = 0.52, β = 2.84 | ---- | ---- |
| 150 nm | α = 0.62, β = 2.44 | ---- | ---- |

**S12.** The full width at half maxima (α) of molecular emission shows the spreading of emission angle (θ) in the back focal plane. The forward to backward gain shows the gain in the directionality of emission. As can be seen α decreases as the thickness of nanowire increases whereas β increases with an increase in the thickness of nanowire. The data confirms that nanowire acts as a reflector for molecular emission from the junction. A thick nanowire will reflect the emission effectively and project it into a narrow range of wavevectors.

## S13. Modes of silver nanowire of thickness 350 nm placed on a glass substrate at an excitation wavelength of 633 nm

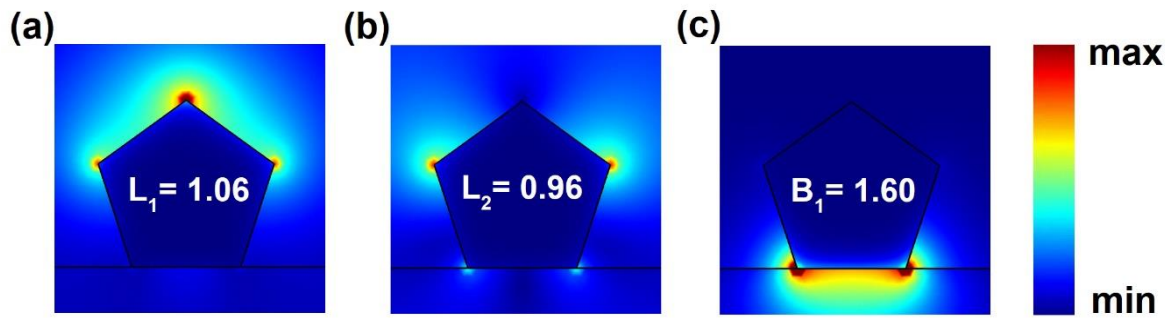

**Figures S13.** Modes of a silver nanowire of thickness 350 nm diameter placed on a glass substrate, at an excitation wavelength of 633 nm [3]. (a) and (b) Near field electric field profile of two leaky modes L1 and L2 with an effective refractive index of 1.06 and 0.96 respectively. (c) Bound mode, B1, with an effective refractive index of 1.60. The leaky modes have a high electric field at the upper vertex of the nanowire, away from the glass substrate whereas the bound mode has a large electric field near the interface of silver nanowire and glass substrate.

**S14. Modes of silver nanowire of thickness 150 nm and 250 nm placed on a glass substrate at an excitation wavelength of 633 nm and 532 nm**

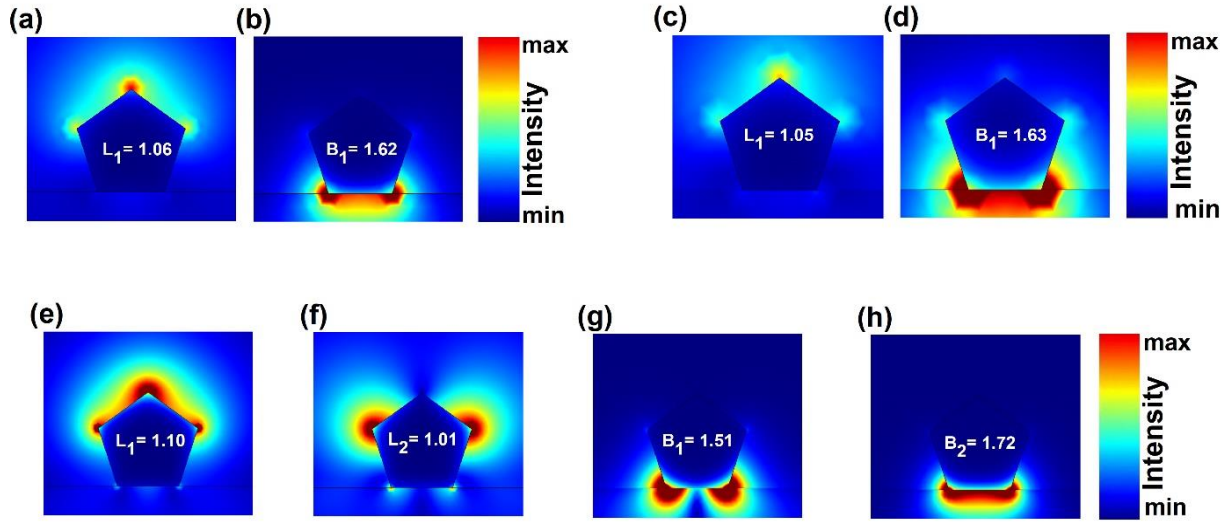

**Figure S14.** Modes of silver nanowire of thickness 150 nm and 250 nm diameter at excitation wavelength of 633 nm and 532 nm.

Figure S14 shows the modes of silver nanowire of thickness 150 nm and 250 nm diameter at an excitation wavelength of 633 nm and 532 nm. (a) − (h) show numerically simulated near field electric field using finite element method.

For an excitation wavelength of 633 nm, (a) and (b) show a leaky mode of effective refractive index 1.06 and a bound mode of effective refractive index 1.62 respectively, for a silver nanowire of thickness 150 nm diameter. c) and d) show a leaky mode of effective refractive index 1.05 and a bound mode of effective refractive index 1.63 respectively, for a silver nanowire of thickness 250 nm diameter. With a nanowire of thickness 250 nm and 150 nm, the number of modes, both leaky and bound, are less as compared to a thicker nanowire of thickness 350 nm.

For a silver nanowire of thickness 350 nm and excitation wavelength of 532 nm, (e)-(h) show two leaky modes, L1 and L2, of effective refractive index 1.10 and 1.01 and two bound modes of refractive index 1.51 and 1.72 respectively.